\documentclass[%
 reprint,
superscriptaddress,
nature
]{revtex4-2}

\usepackage{svg}
\usepackage{xcolor}
\usepackage{graphicx}
\usepackage{dcolumn}
\usepackage{bm}
\usepackage{lipsum}
\usepackage{amsmath}
\usepackage{hyperref}
\usepackage{physics}

\begin{document}

\preprint{}

\title{Free-space quantum information platform on a chip}

\author{Volkan Gurses}
\thanks{Corresponding author: gurses@caltech.edu}
\affiliation{Division of Engineering and Applied Science, California Institute of Technology, Pasadena, CA, USA}
\author{Samantha I. Davis}
\affiliation{Division of Physics, Mathematics and Astronomy, Caltech, Pasadena, CA, USA}
\affiliation{Alliance for Quantum Technologies (AQT), California Institute of Technology, Pasadena, CA, USA}
\author{Neil Sinclair}
\affiliation{Alliance for Quantum Technologies (AQT), California Institute of Technology, Pasadena, CA, USA}
\affiliation{John A. Paulson School of Engineering and Applied Sciences, Harvard University, Cambridge, MA, USA}
\author{Maria Spiropulu}
\affiliation{Division of Physics, Mathematics and Astronomy, Caltech, Pasadena, CA, USA}
\affiliation{Alliance for Quantum Technologies (AQT), California Institute of Technology, Pasadena, CA, USA}
\author{Ali Hajimiri}
\affiliation{Division of Engineering and Applied Science, California Institute of Technology, Pasadena, CA, USA}

\date{\today}

\begin{abstract}

Emerging technologies that employ quantum physics offer fundamental enhancements in information processing tasks, including sensing, communications, and computing \cite{degen2017quantum,gisin2002quantum,Ladd2010}. Here, we introduce the quantum phased array, which generalizes the operating principles of phased arrays \cite{Braun1909} and wavefront engineering \cite{Engheta2006} to quantum fields, and report the first quantum phased array technology demonstration. An integrated photonic-electronic system is used to manipulate free-space quantum information to establish reconfigurable wireless quantum links in a standalone, compact form factor. Such a robust, scalable, and integrated quantum platform can enable broad deployment of quantum technologies with high connectivity, potentially expanding their use cases to real-world applications. We report the first, to our knowledge, free-space-to-chip interface for quantum links, enabled by 32 metamaterial antennas with more than 500,000 sub-wavelength engineered nanophotonic elements over a 550 $\times$ 550 $\mathrm{\mu m^2}$ physical aperture. We implement a 32-channel array of quantum coherent receivers with 30.3 dB shot noise clearance and 90.2 dB common-mode rejection ratio that downconverts the quantum optical information via homodyne detection and processes it coherently in the radio-frequency domain. With our platform, we demonstrate 32-pixel imaging of squeezed light for quantum sensing, reconfigurable free-space links for quantum communications, and proof-of-concept entanglement generation for measurement-based quantum computing. This approach offers targeted, real-time, dynamically-adjustable free-space capabilities to integrated quantum systems that can enable wireless quantum technologies.
\end{abstract}

\maketitle

\section{\label{sec:intro}Introduction}

\begin{figure*}[ht!]
  \includegraphics[width=\textwidth]{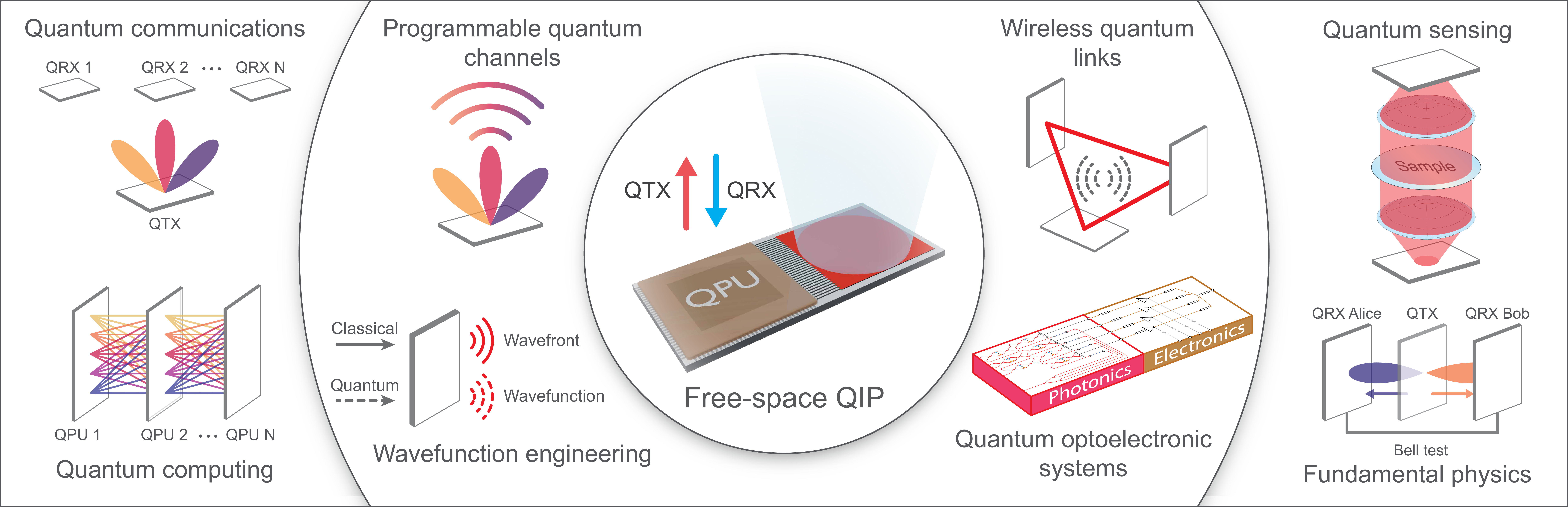}
  \caption{\textbf{Free-space quantum information platform.} Conceptual breakdown of the free-space quantum information platform (QIP) showing the key technology demonstrations and potential applications.}
  \label{fig:fig1}
\end{figure*}

The science and engineering of quantum systems have expanded in the last two decades to enable technologies that can manipulate quantum information at scale \cite{Dowling2003}. These technological developments have led to impressive demonstrations across numerous architectures including superconducting qubits \cite{arute2019quantum}, atom arrays \cite{bluvstein2024logical}, trapped ions \cite{Pino2021}, and integrated photonics \cite{arrazola2021quantum}. Integrated photonics is robust against decoherence at room temperature and varying operating conditions \cite{OBrien2009}. It can be integrated end-to-end with electronics \cite{atabaki2018integrating} and manufactured at high volume and high yield with CMOS fabrication \cite{Margalit2021}. Therefore, it is a compelling candidate for real-world quantum systems including mobile quantum devices. Establishing practical quantum links with high fidelity \cite{yin2012quantum} and a high degree of parallelization \cite{miller2023optics}, offered by free-space interconnects, is important for distributing quantum information between these systems. To our knowledge, there has been no free-space-to-chip interconnect suitable for point-to-point links due to the high coupling loss in conventional free-space-to-chip interfaces and beam divergence challenges in free-space links.

In this work, we demonstrate a large-scale yet compact, room-temperature quantum information platform (QIP), integrated on a silicon photonic chip, which we call the quantum phased array (QPA), that can establish reconfigurable wireless links for free-space quantum information processing. A large-area (550 $\times$ 550 $\mathrm{\mu m^2}$) metamaterial aperture enables a low-loss interface between free space and the chip, and a 32-channel array of self-stabilizing coherent receivers downconverts the quantum optical signals to radio-frequency (RF) by homodyne detection for coherent processing in the RF domain. Coherent processing with the quantum coherent receivers on our QPA chip overcomes the geometric loss limitation \cite{gisin2002quantum}, despite each channel individually exhibiting  high loss. Therefore, we expand the concept of wavefront engineering with classical electromagnetic fields to the non-classical domain. Moreover, coherently interfacing our QPA chip with RF circuits in our platform enables optoelectronic processing of continuous variable quantum information. In this work, we provide a proof-of-concept realization of our platform shown in Fig. \ref{fig:fig1} with the QPA chip. We envision this platform, which allows for free-space connectivity with end-to-end system realization, to enable future mobile and wireless quantum technologies.  

In the following, we outline the QPA system design, demonstrate multipixel squeezed light imaging with 32 coherent channels, introduce wavefunction engineering with quantum phased arrays, and implement a proof-of-concept demonstration of entanglement generation with optoelectronic processing for measurement-based quantum computing.

\section{Quantum phased array chip} \label{sec: QPAchip}
\begin{figure*}[ht!]
  \includegraphics[width=\textwidth]{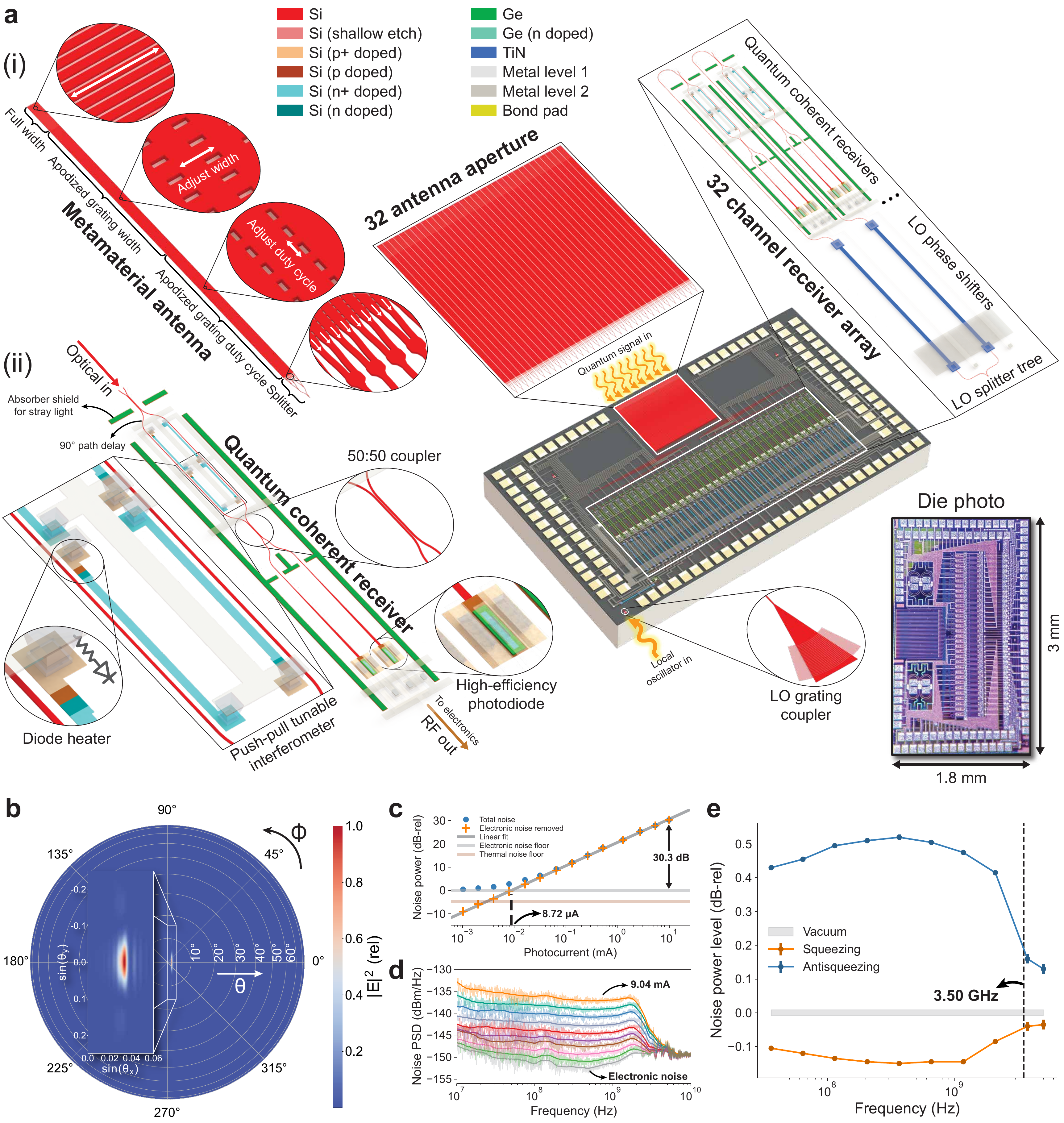}
  \caption{\textbf{Quantum phased array chip}. \textbf{a)} Diagram and die photo (bottom right) of the quantum phased array chip showing the building blocks including i) the metamaterial antenna and ii) the quantum coherent receiver. \textbf{b)} Far-field radiation pattern of the metamaterial antenna. \textbf{c)} LO power sweep characterizing the shot noise clearance and the LO power knee for the QRX in high shot noise clearance configuration. \textbf{d)} Output noise spectra at different LO powers characterizing the shot-noise-limited bandwidth for the QRX in the high bandwidth configuration. \textbf{e)} Squeezed light detection with a single-channel QRX in the high bandwidth configuration.}
  \label{fig:fig2}
\end{figure*}

We implement a QPA receiver system using a commercial silicon photonics process, as shown in Fig. \ref{fig:fig2}.

The QPA chip has 32 channels comprising antennas, waveguides, phase shifters, and quantum coherent receivers. Free-space-to-chip coupling is enabled by a large-area (550 $\times$ 550 $\mathrm{\mu m^2}$) fully-filled aperture comprised of 32 metamaterial antennas (MMAs) that couple a collimated beam in free-space to 32 single-mode waveguides on chip. Each MMA is 597 $\mathrm{\times}$ 16.7 $\mathrm{\mu m^2}$ in footprint. It is designed to have a sufficiently large effective aperture (see Methods) suitable for low-loss coupling with commercial fiber collimators, which are available for beam diameters greater than 200 $\mathrm{\mu m}$. This design minimizes the mode mismatch between the free-space beam and the on-chip aperture to enable low-loss free-space-to-chip coupling, overcoming a limitation in conventional nanophotonic antennas. 

The simulated 3D radiation pattern for the resulting antenna design is shown in Fig. \ref{fig:fig2}b. 
The free-space-to-chip interface is characterized by the geometric loss, which is the loss due to imperfect modal overlap of the incident free-space beam and the on-chip aperture, and the MMA insertion loss, which includes the propagation loss in the MMA and loss due to downward scattering. The MMA has a measured (simulated) insertion loss of 3.82 dB (3.78 dB).  The total measured (simulated) geometric losses of 8 and 32 antenna apertures with a collimated beam of 200 $\mathrm{\mu m}$ diameter, are 2.18 dB (2.03 dB) and 4.85 dB (4.50 dB), respectively. The minimum geometric loss can be attained by adding amplitude weights to the channels, yielding a total measured (simulated) loss of 1.14 dB (1.35 dB). This is at least an order of magnitude lower than previously reported on-chip aperture designs \cite{milanizadeh2022separating, Ashtiani2022} and is sufficiently low to start interfacing free-space quantum optics with photonic integrated circuits (PICs).\par

The waveguides after the antennas are path-length matched and are connected to 32 quantum coherent receivers (QRX). Each receiver consists of a tunable Mach-Zehnder interferometer (MZI), a pair of balanced Ge photodiodes, and a transimpedance amplifier (TIA).
The MZI interferes a signal field with a strong local oscillator (LO) for homodyne detection. The LO is coupled to the chip with a grating coupler and is split into 32 channels with a 1-to-32 splitter tree. The LO input to each channel hosts a thermo-optic phase shifter (TOPS) for phase tuning. Each output of the MZI is sent to a photodiode, and the currents at the outputs of the photodiodes are subtracted and amplified by the TIA.\par

The performance of a QRX is quantified by its shot noise clearance (SNC), LO power knee ($\mathrm{P_{knee}}$), common-mode rejection ratio (CMRR), 3-dB bandwidth ($\mathrm{BW_{3dB}}$) and shot-noise limited bandwidth ($\mathrm{BW_{shot}}$) \cite{Gurses2023, Gurses2022, Bruynsteen2021, Tasker2021}. Our QRXs are characterized in two configurations with two TIA designs: one used in the high bandwidth configuration, optimal for communications, and another used in the high shot noise clearance configuration, optimal for sensing (see Methods). These two TIAs, which trade off bandwidth with noise floor and vice versa, are designed to be interchangeably used with the photonic chip to overcome the fundamental trade-off between SNC and bandwidth in balanced homodyne detection \cite{Gurses2023, Bruynsteen2021}.\par

In the high SNC (high bandwidth) configuration, the QRX has an SNC of 30.3 dB (14.0 dB) and a $\mathrm{P_{knee}}$ of 12.5 $\mathrm{\mu W}$ (521 $\mathrm{\mu W}$). To showcase the high-speed measurements achievable with the QRX in the high bandwidth configuration, squeezed light was measured with a single QRX channel up to the shot-noise-limited bandwidth of 3.50 GHz as seen in Fig, \ref{fig:fig2}d along with the output noise spectra at different LO photocurrents. In both high SNC and high bandwidth configurations, the QRX has a CMRR of 90.2 dB at 1.1 MHz.

In the subsequent experiments, the QPA chip is packaged with electronics on custom printed circuit boards. The QPA chip was wirebonded to an interposer to fan out 104 electronic read-out and control lines to the interfaced electronics. The interposer is plugged into a radio-frequency (RF) motherboard with 50 $\mathrm{\Omega}$ coplanar waveguide outputs. The motherboard hosts a 32-channel TIA array in high SNC configuration and an CMRR auto-correction circuit that is fed back to the MZIs on the PIC (see Methods). The RF outputs are combined with a 32-to-1 power combiner and sent to an RF signal analyzer (ESA). Before power combining, the outputs are also probed with high-impedance outputs for independent recording of the 32-channel signals.

\begin{figure*}[htbp!]
  \includegraphics[width=\textwidth]{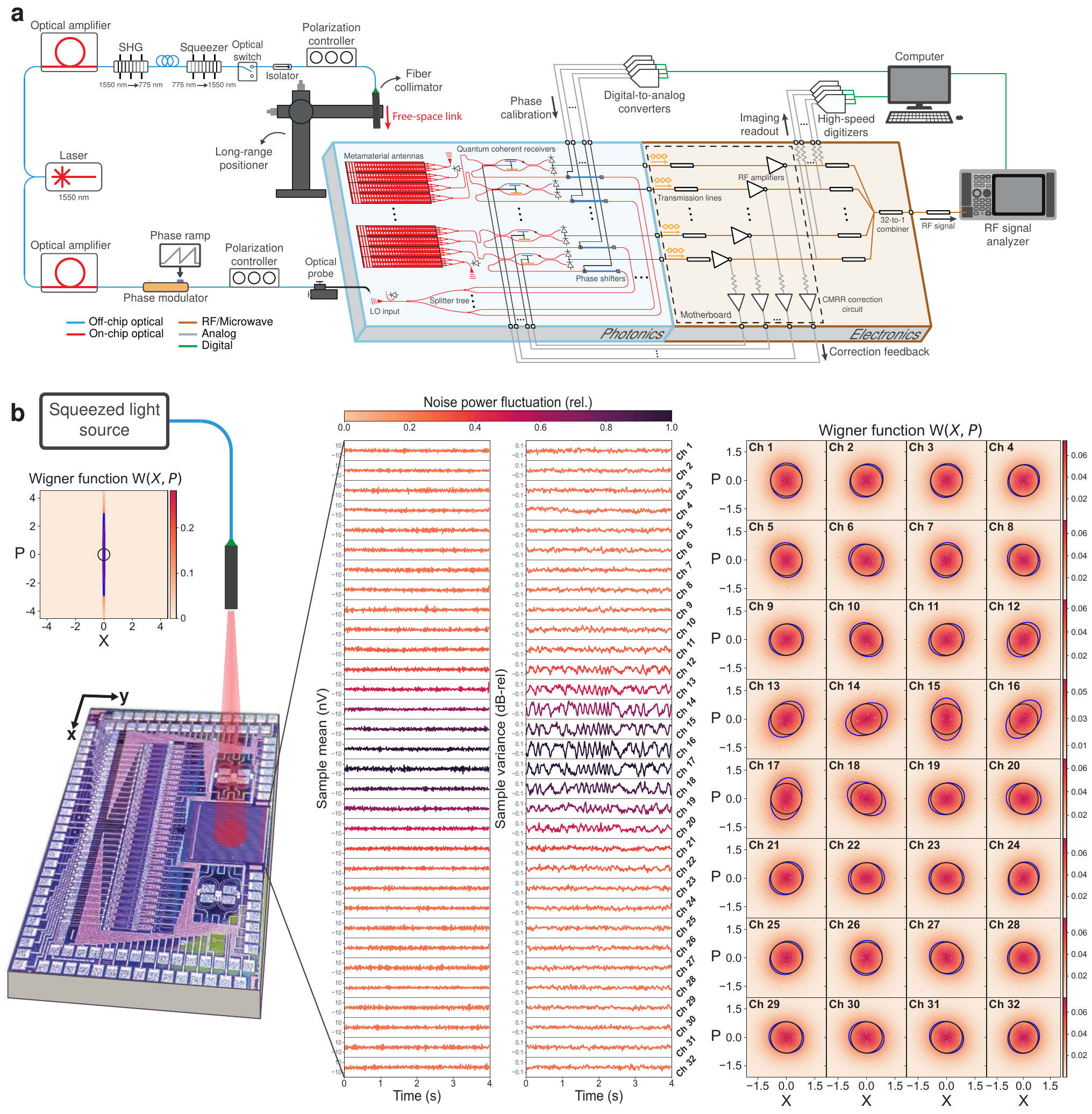}
  \caption{\textbf{Squeezed light imaging.} \textbf{a)} Experimental setup for the squeezed light measurements. Squeezed light is generated off-chip and transmitted over free space to the quantum phased array chip (blue, Photonics), which is interfaced with electronics (orange, Electronics) for quantum information processing. \textbf{b)} Squeezed light imaging measurement.
  Upper left: Wigner function at the source.  The blue and black contours correspond to the half-maximum points of the squeezed vacuum and vacuum states, respectively.
 Middle: The sample mean and variances of the pixel quadratures as a function of time. For each channel, the sample variances are normalized to the mean variance. Right: Wigner functions of the 32 pixel modes. 
}
  \label{fig:fig3}
\end{figure*}
\section{Squeezed light imaging} \label{sec:imaging}

We first operate the QPA chip as a 32-pixel quantum coherent imager. Squeezed light is generated off-chip using off-the-shelf fiber-optic components at a central wavelength of 1550 nm as shown in Fig. \ref{fig:fig3}a.  The squeezed light is sent to a fiber collimator with a 200 $\mathrm{\mu m}$ beam diameter and is transmitted to the chip over free space. At the chip aperture, the collimated squeezed light is spatially distributed across the 32 antennas with a Gaussian amplitude profile. The antennas define a set of 32 pixel modes $\{\hat{a}_j\}$, where $\hat{a}_j$ is the bosonic annihilation operator for the field coupled into the $j$th channel. Each channel outputs a voltage proportional to the phase-dependent quadrature  of its pixel mode, 
\begin{align}
    \hat{X}_j(\theta_j)= \frac{1}{\sqrt{2}}(\hat{a}_je^{i\theta_j}+\hat{a}_j^\dagger e^{-i\theta_j}),
\end{align}
where $\theta_j$ is the phase of the $j$th pixel mode relative to the LO. For a squeezed vacuum field, the quadrature mean $\langle \Delta \hat{X}_j(\theta_j)\rangle$ is zero, and the quadrature variance is,
\begin{align}
    \text{Var}\left(\hat{X}_j(\theta_j)\right) &= \frac{\eta_j}{4} (e^{-2r}\cos^2{\theta_j} + e^{2r}\sin^2{\theta_j}) + \frac{1-\eta_j}{4}, \label{eq:sq_var}
\end{align}
where  $r$ is the squeezing parameter, and $\eta_j$ is the effective efficiency of channel $j$, which includes the effects of source loss, free-space loss, on-chip loss, and RF loss. 

To image the squeezed light, the output voltages are read out to a 32-channel digitizer at a sampling rate of 20 MSa/s. A 0.5 Hz phase ramp is applied to the LO off-chip to acquire voltage samples over various phases, and sample means and variances are calculated over sets of 260,000 voltage samples.
The time evolution of the sample means and variances for all 32 pixel modes are shown in Fig. \ref{fig:fig3}b. The Gaussian profile of the incident field is observed in the amplitude distribution of variances, with the edge channels corresponding to the shot noise power levels for the vacuum state. Without phase locking the signal and LO, phase drifts between the signal and LO paths in fiber optics give rise to additional phase fluctuations. The fluctuations are coherent across all 32 channels due to the preservation of quantum coherence over free space and throughout the chip. For practical applications, phase-locking could be implemented with an optical phase-locked loop \cite{taguchi2022phase} or by co-propagating the LO and signal over free space with the transmitter counterpart to the QPA receiver as illustrated in Fig. 1. 

In Fig. \ref{fig:fig3}b (left), the Wigner function of the source is plotted for the squeezing parameter ($r = 1.95$) as a function of the quadrature observables ($X$, $P$). The Wigner functions of the 32 pixel modes are also plotted in Fig. \ref{fig:fig3}b (right) for the same squeezing parameter as well as the phase and geometric efficiency for each channel. The geometric efficiency of pixel $j$ is the fraction of the incident field that couples onto the $j$th channel. The fixed phase relationships and geometric efficiency (loss) of the channels manifest as variations in the orientation and the squeezing levels of the Wigner functions, respectively (see Methods).

\begin{figure*}[htbp!]
  \centering
    \includegraphics[width=\textwidth]{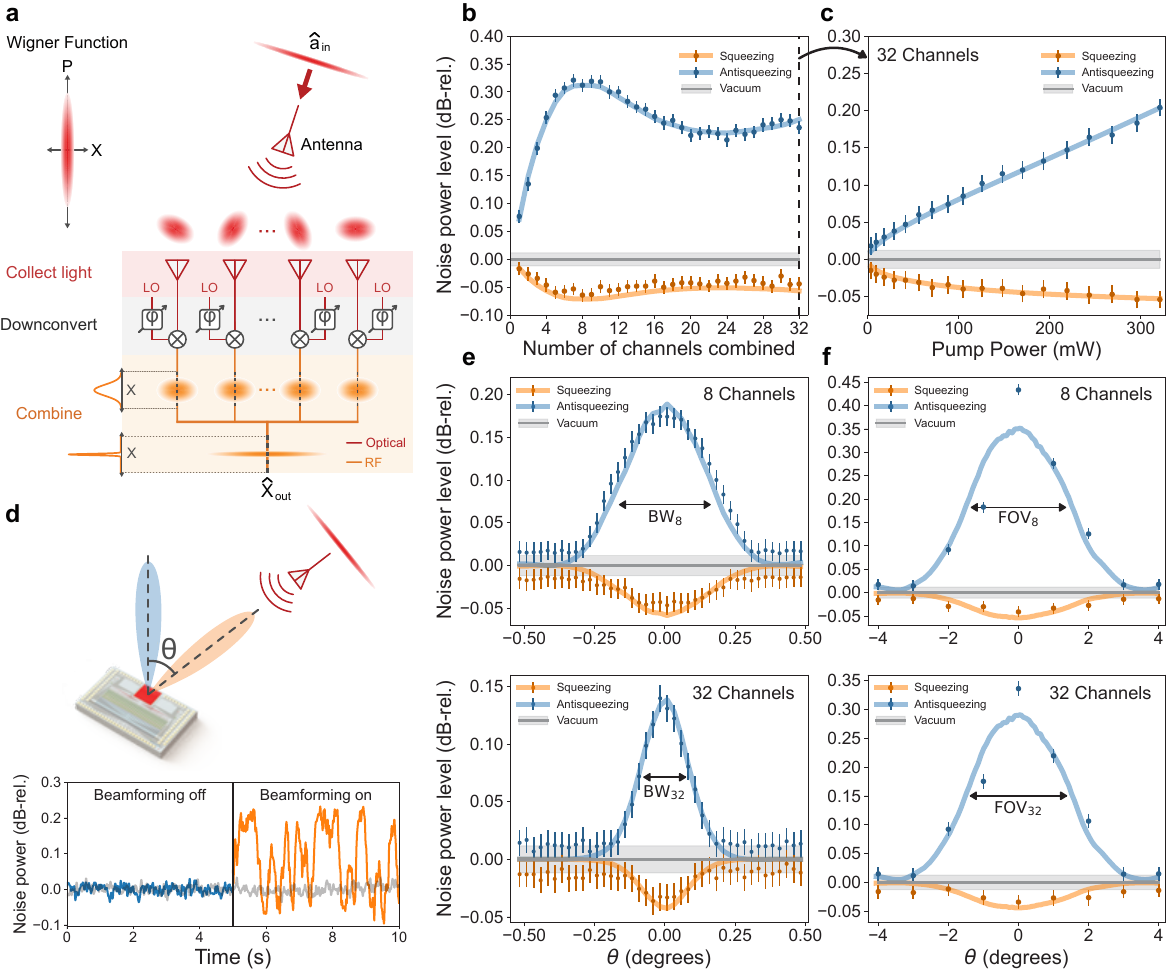}
  \caption{\textbf{Wavefunction engineering.} \textbf{a)} Conceptual illustration of wavefunction engineering with squeezed light. \textbf{b)} Squeezing (orange) and antisqueezing (blue) levels as a function of the number of channels combined relative to the vacuum level (grey) after the chip is beamformed at the squeezed light transmitter. \textbf{c)} Source characterization showing squeezing and antisqueezing levels as a function of squeezer pump power with 32 channels combined. \textbf{d)} Demonstration of programmable and spatially selective free-space quantum links, showing the lack of quantum signal when the receiver is beamformed at empty space (blue) and the reception of a quantum signal when the receiver is beamformed at the transmitter (orange). The grey trace is the vacuum signal. \textbf{e)} Squeezing and antisqueezing levels characterizing the beamwidth of the quantum link for 8 (top) and 32 (bottom) channels combined. \textbf{f)} Squeezing and antisqueezing levels characterizing the field of view of the QPA receiver for 8 (top) and 32 (bottom) channels combined.
 }
  \label{fig:fig4}
\end{figure*}
\begin{figure*}[htbp!]
  \centering
    \includegraphics[width=\textwidth]{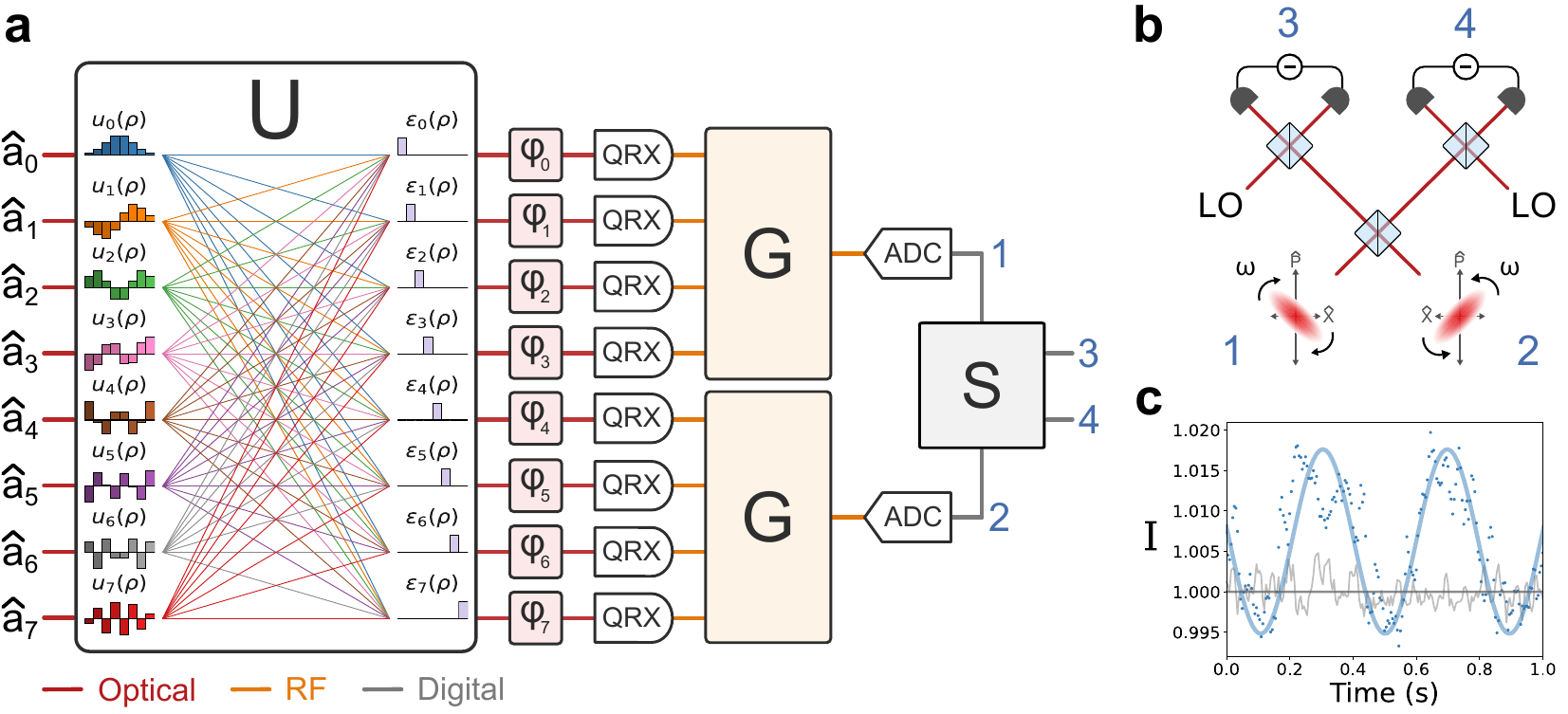}
  \caption{\textbf{Quantum optoelectronic processing.} 
 \textbf{a)} Quantum circuit architecture for entanglement generation. The free space operation ($\mathrm{U}$) corresponds to the change-of-basis matrix mapping the spatial modes of the input state to the pixel modes, where $\rho$ represents the spatial coordinates in the aperture plane of the chip. Each colored line represents a matrix element corresponding to the overlap of an antenna and pixel mode function. A phase shifter $\varphi_j$ is applied to each pixel mode, and each half of the array is combined in a 16:1 RF power combiner ($\mathrm{G}$). The output voltages of the power combiners are digitized and followed by a beamsplitter transformation ($S$). 
  \textbf{b)} Emulated optical circuit for two-mode Gaussian cluster state generation.
  \textbf{c)} The cluster state inseparability (I) measured over time for a linear phase ramp. The data for the squeezed vacuum and vacuum states are plotted in blue and gray, respectively. The solid lines are the analytical expectations with a sinusoidal fit to the squeezed data. 
  }
  \label{fig:fig5}
\end{figure*}
\section{Wavefunction engineering} \label{sec:wavefunction_engineering}

Geometric loss is a longstanding challenge in quantum communications \cite{gisin2002quantum}. In a point-to-point free-space quantum link, a transmitter encodes a quantum signal in a beam of light that is sent to a receiver. The spot size of the beam spreads with distance due to diffraction, resulting in geometric loss from the overlap of the diverging spot size and the receiver aperture area. Diffraction-induced geometric loss can result in severe signal loss that ultimately limits the range and rate of quantum communication protocols \cite{pirandola2021limits,krvzivc2023towards,liao2017satellite,ren2017ground}. The effect of geometric loss is apparent in Wigner functions of  Fig. \ref{fig:fig3}b, where the degree of squeezing per channel is reduced compared to that of the transmitted state. In classical wireless communications and sensing, beam divergence is controlled by wavefront engineering with transmitter or receiver phased arrays. Wavefront engineering allows for active manipulation of an electromagnetic field in a dynamic real-time fashion \cite{Hansen2009,Li2018}. Beamforming, or angular focusing, of an electromagnetic field is performed by coherently combining elements in a phased array such that the signal field constructively interferes at a selected angle \cite{van1988beamforming}. Here, we extend beamforming to quantum fields and demonstrate how wavefunction engineering with quantum phased arrays can overcome geometric loss, enable spatial selectivity, and establish reconfigurable and programmable quantum links.\par

Beamforming with a QPA receiver is illustrated in Fig. \ref{fig:fig4}a. A quantized electromagnetic field with annihilation operator $\hat{a}_\text{in}(\rho)$ is transmitted to a phased array of quantum coherent elements. The field $\hat{a}_\text{in}(\rho) =\sum_n u_n(\rho)\hat{a}_n$ is expanded over a set of independent modes with annihilation operators $\{\hat{a}_n\}$, where $\rho$ represents the transverse coordinates in the plane orthogonal to propagation. The modes have an associated set of orthonormal mode functions $\{u_n(\rho)\}$ that correspond to photon-wavefunctions in second quantization \cite{cohen1989photons, mandel1995optical, grynberg2010introduction}.  Due to diffraction, a portion of the field is coupled onto each antenna, resulting in high geometric loss per channel. The RF outputs of the coherent receivers are combined after applying a gain and phase to each channel. In the large array limit, the combined RF output is proportional to the quadrature of an output field, 
\begin{align}
    \hat{a}_\text{out}(\rho) = \sum_n \hat{a}_n \int g(\rho)e^{i\phi(\rho)} u_n(\rho) d\rho,
\end{align}
where $g(\rho)$ and $\phi(\rho)$ are the applied gain and phase profiles (see Methods). The gain and phase profiles give rise to a programmable array mode function $\mathcal{A}(\rho)=g(\rho)e^{i\phi(\rho)}$, which is used to engineer the wavefunction of the input field. For a signal encoded in a mode $\hat{a}_n$, perfect modal overlap, i.e. geometric efficiency, is achieved by setting $\mathcal{A}(\rho) = u_n(\rho)$. Beamforming refers to the optimization of the gain and phase profiles to maximize the geometric efficiency. For the QPA chip, the minimum geometric loss of 1.14 dB for a 200 $\mu$m beam diameter sets an upper bound of 0.77 on the geometric efficiency, which can approach unity in future chip iterations with specialized antenna design and by scaling up the array.

By beamforming with the QPA chip, we experimentally overcome the geometric losses of Fig. \ref{fig:fig3}. 
After optimizing the LO phases for all 32 channels (see Methods), squeezed light is transmitted to the chip through the fiber collimator. A 1-Hz phase ramp is applied to the LO before coupling to the chip, and the outputs of the channels are coherently combined with a 32:1 RF power combiner. The combined output signal is sent to an RF signal analyzer, which measures the noise power proportional to the quadrature variance in Eq. \ref{eq:sq_var}. 

The improvement in geometric efficiency with beamforming is demonstrated in Fig. \ref{fig:fig4}b. The noise powers are measured for various number of combined channels.
The number of channels is increased symmetrically about the center of the array, starting with only channel 16 connected to the power combiner. The solid lines are a fit of the data to a model constructed from the classically characterized signal-to-noise ratio for each channel combination. The squeezing (antisqueezing) level improves from $-0.017 (+0.077)\pm 0.012$ dB for a single channel to $-0.064(+0.312) \pm 0.012$ dB for eight channels combined, corresponding to an increase in the geometric efficiency by a factor of $4.5$ (see Methods). 

We perform a source characterization to confirm that the combined RF signals correspond to squeezed light with increased efficiency.
For all 32 channels combined, the squeezing and antisqueezing levels for various squeezer pump powers ($P$) are shown in Fig \ref{fig:fig4}c. The solid lines correspond to a least-squares fit of the data to a model, where the effective efficiency of the system, $\eta$, and the spontaneous parametric down conversion (SPDC) efficiency, $\mu$, are taken as floating parameters. 
 We obtain $\eta = 0.016$ and $\mu = 0.038$ $[\text{mW}]^{-1/2}$, which is consistent with the SPDC efficiency of the squeezer (see Methods).\par

Establishing spatially-selective free-space quantum links with our QPA is illustrated in Fig. \ref{fig:fig4}d. A quantum link is established by beamforming the QPA chip at an angle at which no light is detected (blue). The phase settings are reconfigured by applying a linear phase mask to the LO phase shifters, which steers the link towards a squeezed light source (orange). In the first five seconds, no quantum signal is observed, demonstrating successful spatial filtering. In the next five seconds, after the quantum link is electronically steered toward the source, the noise power modulations of the squeezed light are observed. \par

The spatial selectivity, namely beamwidth, of a quantum link is characterized in Fig. \ref{fig:fig4}e. A classical phase calibration is performed for a beam of squeezed light at a normal angle of incidence to the chip. The angle of incidence ($\theta$) of the beam is swept while the chip is kept beamformed to normal incidence ($\theta = 0$), for 8 and 32 channels combined. The squeezing and antisqueezing levels as a function of $\theta$ are shown in Fig. \ref{fig:fig4}e. The solid lines are fits of the data to a model obtained from classical beamwidth measurements. The beamwidths ($\mathrm{BW_n}$) for 3-dB loss are 0.41 degrees and 0.20 degrees with 8 channels and 32 channels combined, respectively, demonstrating the increased spatial selectivity as the array is scaled (see Methods).

The programmability of free-space quantum links over the field-of-view (FoV) of the QPA chip is demonstrated in Fig. \ref{fig:fig4}f. At each of the nine different angles ($\mathrm{\theta}$), classical phase calibration is performed and the optimal phase settings are recorded. The squeezed light is then transmitted to the chip at each of the nine angles. For each angle, a quantum link is programmed by applying the corresponding phase settings to the LO phase shifters. The squeezing and antisqueezing levels for each angle are shown in Fig. \ref{fig:fig4}f for 8 and 32 channels combined. The solid lines are a fit to a model obtained from classical FoV measurements. The 3-dB loss FoV for squeezed light is 2.3 degrees and 2.7 degrees with 8 channels and 32 channels combined, respectively (see Methods).

\section{Quantum optoelectronic processing} \label{sec:qoe}

We perform a proof-of-concept demonstration of cluster state generation in a measurement-based approach \cite{armstrong2012programmable,ferrini2013compact} to illustrate the potential of quantum optoelectronic systems.
Cluster states are a class of entangled graph states that form a resource for universal measurement-based quantum computation \cite{raussendorf2001one}. CV Gaussian cluster states can be generated by interfering squeezed states in linear optical networks \cite{van2007building, yukawa2008experimental}. 
Here we generate two-mode cluster state correlations by implementing the equivalent linear optical network after optoelectronic downconversion with an RF circuit. \par

The quantum circuit architecture is shown in Fig. \ref{fig:fig5}a. A squeezed state is transmitted over free space to the QPA chip and a phase ramp at a modulation frequency of $f = 0.5$ Hz is applied to the LO before coupling to the chip. The RF outputs of the QRXs in each half of the array are sent to a 16:1 power combiner. Beamforming is performed on all 32 channels such that the two outputs of the power combiners are in phase. To improve the geometric efficiency, the outermost 12 channels are disconnected from each 16:1 power combiner, for a total of 8 pixel modes in Fig. \ref{fig:fig5}a.
The outputs of the power combiners are digitized at a sampling rate of 100 MSa/s and an RF beamsplitter transformation is emulated on the digitized quadratures (see Methods).

The inseparability criterion required to show cluster state entanglement is,
\begin{align}
    I = \text{Var}(\hat{P}_4- \hat{X}_3) + \text{Var}(\hat{P}_3- \hat{X}_4) < 1 \label{eq:insep}
\end{align}
where $\hat{X}_i, \hat{P}_i$ are the quadrature operators for each cluster state mode denoted by $i=3,4$ in Fig. \ref{fig:fig5}b, and the variances are relative to those of the vacuum state. The quadrature correlations $I$ as a function of time are shown in Fig. \ref{fig:fig5}b. We observe the sinusoidal signature expected for a rotation of the measurement basis due to the LO phase modulation. We obtain a minimum inseparability of $I= 0.994 \pm 0.002$, which violates the classical bound by three standard deviations. The resolution of entanglement is enabled by the high precision and stability offered by the chip-scale optoelectronics (see Methods).

\section{Discussion and outlook}
We demonstrate a compact, scalable free-space quantum information platform integrated on a silicon photonic chip with more than 1000 functional components on a 3 $\mathrm{\times}$ 1.8 $\mathrm{mm^2}$ footprint. We design and implement a sub-wavelength engineered large active area metamaterial aperture with more than 500,000 scattering elements and a 32-channel array of state-of-the-art quantum coherent receivers. 
To our knowledge, we report the first free-space coupling and processing of non-classical states to a system-on-chip, providing an interface between free-space quantum optics and integrated photonic systems. We expand classical wavefront engineering to quantized electromagnetic fields and demonstrate wavefunction engineering with quantum phased arrays that enable dynamically programmable, wireless quantum links with spatial selectivity.  With our integrated platform, we mitigate the long-standing challenge of diffraction-induced geometric loss by coherently combining the signals of the receiver outputs in RF thus enabling wireless quantum communications. 
We realize a measurement-based entanglement generation scheme with quantum optoelectronic processing by implementing operations on downconverted quantum optical information with RF circuits.

Demonstrated improvements in component losses \cite{vahlbruch2016detection,Michel2010,Boes2023} chart a path toward deployment of our platform to real-world applications including microscopes, sensors \cite{Rogers2021}, and quantum communication transceivers, as well as potential investigations of fundamental physics research \cite{tse2019quantum}. Interfacing integrated quantum photonics with electronics in the same package enables novel engineering opportunities in realizing large-scale room-temperature quantum systems. Coherent processing of downconverted quantum optical information with RF or microwave integrated circuits could enable a low-loss optoelectronic approach to quantum information processing as a quantum analog of microwave photonics \cite{Marpaung2019}.

Our novel quantum platform bridges integrated photonics, electronics, and free-space quantum optics with multiple envisioned applications in fundamental physics and engineering. 

\subsection{Acknowledgements}
We are grateful to Raju Valivarthi for technical support on the squeezed light sources and for technical discussions, Pablo Backer Peral for aiding the development of digitizer script, Esme Knabe for aiding the initial single-channel QRX measurements, Debjit Sarkar for aiding the setup for the high bandwidth squeezed light measurement, and Andrew Mueller for aiding figure design. Support for this work was provided in part by the Carver Mead New Adventures Fund and Alliance for Quantum Technologies’ (AQT) Intelligent Quantum Networks and Technologies (INQNET) program. S.I.D. is in part supported by the Brinson Foundation. M.S. is in part supported by the Department of Energy under Grant No. SC0019219.

\bibliography{references.bib}

\section{Methods}
   
\subsection{Theory}
Consider a quantized electromagnetic field $\hat{E}$ propagating over free space in the $z$ direction towards a quantum phased array receiver. The field is normally incident to the aperture plane at $z=0$, and can be decomposed into positive and negative frequency components, $\hat{E} = \hat{E}^{+} + \hat{E}^{-}$, where $\hat{E}^{\pm}$ are complex conjugates. At the aperture plane, the positive frequency component of the field can be expressed as,
\begin{align}
    \hat{E}^{+}(\rho,t) = \sqrt{\frac{2\pi \hbar \omega}{L}}\sum_n \hat{a}_n u_n(\rho)e^{i\omega t}, \label{eq:qed}
\end{align}
where $\rho$ represents the transverse coordinates $(x,y)$, $\omega$ is the frequency, and $L$ is the quantization length \cite{beck2000quantum}. The field is expanded over a complete set of independent modes with bosonic annihilation operators $\{\hat{a}_n\}$ satisfying $[\hat{a}_n,\hat{a}_m^\dagger] = \delta_{nm}$. Relevant to fields in free space are the Hermite-Gaussian modes and associated mode functions. In Eq. \ref{eq:qed}, we assume a monochromatic treatment of the field. We note that the squeezed light generated by SPDC in the experiments is broadband and that the following analysis can be extended to multiple spectral modes.

The incident field to the aperture can be represented with the annihilation operator,

\begin{align}
    \hat{a}_{\text{in}}(\rho) = \sum_n \hat{a}_n u_n(\rho).
\end{align}

The field coupled onto the $j$th channel of the receiver corresponds to the modal overlap of the incident field and the $j$th antenna,
\begin{align}
    \hat{a}_j &= \int \mathcal{E}_j(\rho) \hat{a}_\text{in}(\rho) d\rho =   \sum_n U_{jn} \hat{a}_n \label{eq:pixel_field_mode}
\end{align}
where $\hat{a}_j$ is the creation operator for the $j$th pixel mode and $\mathcal{E}_j(x) $ is the mode function for the $j$th antenna. The coupling corresponds to a change-of-basis transformation $U$ between the input and pixel modes, 
\begin{align}
    U_{jn} = \int \mathcal{E}_j(\rho)  u_n(\rho) d\rho.
\end{align}
For a signal encoded in a particular mode of a field with all other modes in the vacuum state, imperfect overlap of antenna and signal mode functions causes spurious vacuum modes to couple onto an antenna, resulting in geometric loss.

To correct for geometric loss, the outputs of the receivers are combined in RF after applying a gain and phase shift to the output of each receiver. The output signal measured with the RF signal analyzer is proportional to the quadrature, 
\begin{align}
    &\hat{X}_\text{out}= \frac{1}{\sqrt{2}}(\hat{a}_\text{out}e^{i(\omega - \omega_{LO}) t}+\hat{a}_\text{out}^\dagger e^{-i(\omega - \omega_{LO})  t}), \label{eq:xout}
\end{align}
 where $\omega_\text{LO}$ is the frequency of the local oscillator, and the downconverted frequency $\omega - \omega_{LO}$ is RF. The quadrature $\hat{X}_\text{out}$ corresponds to an output field,
\begin{align}
    \hat{a}_{\text{out}}  = \sum_j g_j e^{i\phi_j}\hat{a}_j \approx \sum_n \hat{a}_n \int \mathcal{A}(\rho) u_n(\rho) d\rho, \label{eq:aout}
\end{align}
where $g_j$ is the gain and $\phi_j$ is the net phase applied to output $j$. The approximation is taken in the limit of an array with a large number of narrow pixels, where the gains and phase shifts approach continuous amplitude and phase distributions $g(\rho)$ and $\phi(\rho)$, respectively. In Eq. (\ref{eq:aout}), $\mathcal{A}(\rho)=g(\rho)e^{i\phi(\rho)}$ is a programmable array mode function that enables wavefunction engineering. For a signal encoded in a mode $\hat{a}_n$, perfect modal overlap can be achieved by setting $\mathcal{A}(\rho) = u_n(\rho)$ through beamforming. Due to the orthonormality of $\{u_n(\rho)\}$, the vacuum noise contributions across all the receivers destructively interfere, resulting in unity geometric efficiency. For multimode fields, the signal in a particular mode can be uniquely selected by setting $\mathcal{A}(\rho)$ to the desired mode function \cite{beck2000quantum, dawes2003mode, fabre2020modes}. Furthermore, the QPA acts a tunable spatial filter, rejecting quantum signals from angles that result in destructive RF interference for a given phase setting. This defines an angular ``beamwidth" for a quantum link established by beamforming. 

The transformations on the input modes can be extended to matrix operations after coherent detection \cite{ferrini2013compact}. 
The overall class of operations that can be performed by the QPA chip are, 
\begin{align}
    \vec{a}_\text{out} = M D U \vec{a}_\text{in},
\end{align}
where the input modes are grouped into the vector $\vec{a}_\text{in} = (\hat{a}_1,\hat{a}_2,...)$, $U$ is the change-of-basis unitary, $D = \text{diag}(g_1 e^{i\phi_1},g_2 e^{i\phi_2},...)$, and $M$ is a matrix implemented after coherent detection.
\subsection{Chip design}
Decoherence due to loss is one of the biggest challenges for integrated quantum photonics \cite{OBrien2009}. The most significant source of loss for free-space-coupled integrated systems is geometric loss due to the mode mismatch between an impinging beam mode and the aperture mode. In the case of a collimated beam such as a beam transmitted from a large-aperture transmitter, beam divergence due to diffraction is the primary cause of mode mismatch between the impinging beam and the receiver aperture. Reducing this mode mismatch requires the QPA receiver to have a large enough effective aperture. The effective aperture can be increased either by arraying a large number of small-area antennas or employing a single large-area antenna. We demonstrate both approaches in the design by arraying 32 large-area antennas. Due to the planar routing constraints and the resulting loss from having feed waveguides inside the partially filled aperture, we demonstrate a 1D array. Multi-layer apertures can be used to expand this concept into 2D arrays \cite{Ashtiani2021}.\par
To maximize the effective aperture of a single antenna, the coupling strength per area needs to be minimized. To achieve this, various antenna topologies were simulated, and a parallelized waveguide metamaterial antenna design was determined to have the lowest scattering strength per area while abiding by the foundry design rules. Sixteen waveguides were connected and parallelized with sub-wavelength gratings in the regions between the waveguides. The 0.82 $\mathrm{\mu m}$ wide waveguides keep a single mode confined throughout the length of the antenna so that the phasefront of the coupled light across the cross-section of the antenna is flat. At the end of the antenna active area, a mode converter comprising a taper couples the light from 0.82 $\mathrm{\mu m}$ waveguides to 0.5 $\mathrm{\mu m}$ waveguides. A Y-junction-based 16-to-1 combiner tree combines all the outputs from a single antenna into a single mode propagating in the 0.5 $\mathrm{\mu m}$ wide waveguide that is used to route the quantum signal on the PIC. Three grating regions with apodized coupling strengths to mode match the amplitude profile of the impinging beam were designed, as seen in Fig. 2. The antenna design was verified using an FDTD simulation. The physical footprint of the antenna is 597 $\times$ 16.7 $\mathrm{\mu m^2}$. Across 597 $\mathrm{\mu m}$ length starting from the splitter tree, the 0 $\mathrm{\mu m}$ to 47 $\mathrm{\mu m}$ is the splitter tree, 47 $\mathrm{\mu m}$ to 347 $\mathrm{\mu m}$ is the apodized grating duty cycle region, 347 $\mathrm{\mu m}$ to 547 $\mathrm{\mu m}$ is the apodized grating width region, and 547 $\mathrm{\mu m}$ to 597 $\mathrm{\mu m}$ is the full width region. The aperture of the chip comprises 32 of these antennas with 17.5 $\mathrm{\mu m}$ pitch to ensure sufficiently low crosstalk between antennas. To aid with the free-space alignment of an impinging field to the chip and ensure uniform response across all 32 antennas, two antennas are added on each side of the aperture, resulting in 36 total antennas. One antenna on each side is connected to a standard grating coupler to aid alignment with optical input/output, and the other antenna is connected to a photodiode to aid alignment with optical input/electronic output. \par
The QRX design comprises a tunable Mach-Zehnder interferometer (MZI) made out of two 50:50 directional couplers and two diode phase shifters. Each phase shifter is 100 $\mathrm{\mu m}$ long, comprising a resistive heater made out of doped silicon with 1 $\mathrm{\Omega}$ resistance and a diode in series with 1 V forward voltage. Doped Si is placed 0.9 $\mathrm{\mu m}$ away from the waveguides to minimize loss from free carriers. The MZI is configured in a push-pull configuration to extend the tuning range of the coupling coefficients and is designed to provide sufficient tuning with $\mathrm{\pm}$5 V drivers. One branch of the MZI includes an optical delay with 90$^\circ$ phase shift to set the nominal coupling of the MZI to 50:50. Fabrication imperfections such as changes in the gap in the coupling region of the couplers and surface roughness in the waveguides between the couplers shift this ideal 50:50 coupling randomly throughout the chip. The tunability of the MZIs allows correcting for these imperfections to set 50:50 coupling. The MZIs are also designed to be symmetric to ensure a high extinction ratio.\par
After the MZI, the waveguides are adiabatically tapered to connect to a balanced Ge photodiode pair with \textgreater20 GHz bandwidth at 3 V reverse bias, \textgreater70$\%$ quantum efficiency, and \textless100 nA dark current. The QRX is surrounded by a Ge shield to absorb stray light propagating in the chip substrate and prevent it from coupling to the photodiodes. Each QRX output is connected to a separate on-chip pad to be interfaced with a transimpedance amplifier (TIA) and subsequent electronics for RF processing.\par
The LO is coupled to chip with a standard grating coupler and is sent to each QRX through a 1-to-32 splitter tree. Each Y-junction in the splitter tree has 0.28 dB loss, and the grating coupler has 3.30 dB loss. Before the splitter tree, a directional coupler on the LO waveguide is present to couple 1\% of LO power to a monitor photodiode for LO power monitoring. After the splitter tree, a TOPS is included in each branch to tune the quadrature phase of each channel for the phase calibration of the system. Each TOPS for phase tuning is 315 $\mathrm{\mu m}$ long, comprising a resistive heater made out of titanium nitride above the waveguide with 630 $\Omega$ resistance. 

\subsection{Chip fabrication}
The QPA PIC was fabricated in the AMF 193-nm silicon-on-insulator (SOI) process. The process has two metal layers (2000-nm thick and 750-nm thick) for electronic routing, a titanium nitride heater layer, a 220-nm thick silicon layer, a 400-nm thick silicon nitride layer, germanium epitaxy, and various implantations for active devices. A process design kit (PDK) from the foundry was provided, and the final design was completed and verified using the KLayout software.

\subsection{Squeezed light generation}
To generate squeezed light, continuous wave light from a fiber-coupled 1550 nm laser is split into a signal path and a local oscillator (LO) path. The light in each path is amplified by an erbium-doped fiber amplifier. After amplification in the signal path, the 1550 nm coherent light is upconverted to 775 nm by a periodically poled lithium niobate (PPLN) waveguide via second harmonic generation. The upconverted light is used as a continuous-wave pump for Type 0 spontaneous parametric down-conversion (SPDC) with another PPLN waveguide, which generates broadband light in a squeezed vacuum state at a central wavelength of 1550 nm. 
A total of four PPLN waveguides were used to generate squeezed light in the experiments. The squeezed vacuum light is sent to a fiber optic collimator, which transmits the light over free space with a flat phase front to the chip aperture. After amplification in the LO path, the 1550 nm coherent light is sent to a bulk lithium niobate electro-optic modulator for phase control. The phase-modulated local oscillator is sent to a cleaved fiber which is grating-coupled to the LO input of the chip. Polarization controllers before the collimator and on the LO fiber are used to optimize coupling efficiency to the chip.

\subsection{System electronics}
The QPA chip is first packaged with an interposer board for fanning the electronic input/output (IO) to/from the chip. The interposer board is designed with a laser-milled cavity in the middle to place the QPA chip surrounded by pads with blind vias for high-density routing. The chip and the interposer are assembled so that the on-chip pads are level and parallel with the on-board pads to shorten the bond wire length. The traces from the interposer pads to the TIA inputs on the motherboard are minimized and spaced sufficiently apart to minimize electronic crosstalk with 50 $\Omega$ coplanar waveguide (CPW) transmission lines. The discrete TIA circuit on the motherboard utilizes a FET-input operational amplifier (op-amp) with resistive feedback. The op-amp IC (LTC6269-10) has a 4 GHz gain-bandwidth product and is used with a 50 $\mathrm{k\Omega}$ feedback resistor. The capacitance of the feedback trace is used to ensure sufficient phase margin while keeping the closed-loop gain greater than 10 since the op-amp is decompensated. A 50 $\mathrm{\Omega}$ resistor is placed in series with the output of the TIA for impedance matching and to dampen any oscillations from capacitive loading at the output. The TIA outputs are routed with 50 $\Omega$ CPW transmission lines to a high-speed, high-density connector to route the signals to data acquisition.\par

The DC voltage across the TIA feedback resistor is used as the error signal for the CMRR correction and drives an integrator circuit with a chopper-stabilized op-amp IC (OPA2187) for low voltage offset, flicker noise, and offset drift. The integrator unity-gain bandwidth is set close to DC to dampen any oscillations in the CMRR auto-correction feedback. The integrator's output is fed back to the MZI on the QPA chip to correct the CMRR continuously. The polarity of the integrator is designed to match with the polarity of the push-pull MZI so that the correction circuit always maximizes the CMRR whether the imperfections lead to negative or positive DC current from the balanced photodiodes. The correction is limited by the dark current of each QRX and the offset voltage at the input of each integrator, but offset correction can be applied to each integrator to further maximize the CMRR. The CMRR auto-correction circuit extracts an error signal from the TIA output, probing the imperfect CMRR of each QRX and feeding it back to each respective push-pull MZI to continuously correct the CMRR of the QRX array. This ensures shot-noise limited noise floor during chip measurements, maximizing the shot noise clearance and effective efficiency. A high-speed coaxial cable assembly is used to connect to the high-density connector on the motherboard. The cable first connects to a power board powering the active electronics on the motherboard. This board also routes the output from the two photodiodes connected to the two edge antennas of the aperture and the output from the monitor photodiode connected to the LO coupler to current meters for continuous monitoring of signal and LO alignment on the chip. Another cable then connects the remaining IO to a splitter board that splits the 32 QRX outputs for simultaneous imaging and RF data acquisition. The remaining control lines for tuning the on-chip TOPS are connected to 32 digital-to-analog converters (DACs) for independent phase tuning of each QRX.

\subsection{Data acquisition and readout}
The 32 QRX outputs after the splitter is connected to boards that host SMA connectors to interface with data acquisition (DAQ) equipment. One board, used for parallelized 32-channel readout, connects to 32 channels of digitizers with 100 MHz bandwidth, 100 MSa/s adjustable sampling rate and 14-bit resolution. The digitizers are used in high-impedance mode to read out the voltage of each QRX output for squeezed light imaging and during RF measurements. The other board, used for RF single channel readout, connects to a 32-to-1 RF power combiner assembly with an operating frequency range of 0.1-200 MHz. The output from the power combiner is connected to the ESA. For squeezed light measurements in Fig. \ref{fig:fig4}b,c,e and f, the ESA is configured to be used in the zero-span mode at a center frequency of 5.5 MHz, with a resolution bandwidth of 2 MHz, and a video bandwidth of 5 Hz. For the measurements in Fig. \ref{fig:fig4}, the video bandwidth is 10 Hz. Center frequency and resolution bandwidth are selected to maximize the shot noise clearance after doing a parameter sweep.

\subsection{Phase calibration}
For each angle of incidence, we optimize the settings for the 32 LO TOPS such that the quadratures for all pixel modes are aligned to the same phase. Precise phase calibration is crucial to prevent additional loss due to vacuum noise leaking into the combined output. Phase calibration is performed with a 1550 nm coherent state transmitted by the collimator and a 5 MHz phase ramp is applied to the LO before coupling to the chip. The 5 MHz downconverted RF signal after the QRX outputs are combined is used as feedback to the computer to tune the on-chip TOPS iteratively. Various signal processing schemes and algorithms have been developed for beamforming in classical phased arrays such as random search, gradient search, direct matrix inversion, and recursive algorithms \cite{Monzingo2011}. We employ a modified gradient search algorithm by sweeping phase settings of on-chip TOPS with an orthogonal mask set.  We sweep the TOPS voltage starting with large voltage steps and continuing with progressively smaller voltage steps with each optimization iteration. Due to the Gaussian amplitude front, edge channels contribute less SNR to the combined output. Therefore, we sweep channel settings starting from the edge channels and continuing to the middle channels. As each channel is tuned and the total SNR improves, the proportional increase in SNR from element to element gets smaller, leading to higher errors in the optimal phase setting of the last channels that are swept. Therefore, for each optimization iteration, we reverse the order of channels to be swept.

\subsection{Data analysis}
The squeezing and antisqueezing levels are obtained from a statistical analysis of the quadrature sample variances or noise powers. For the experiments in Fig. 3(4,5), quadrature sample variances (noise powers) are acquired for squeezed vacuum and vacuum states over an approximately uniform distribution of phases, and histograms are constructed for the acquired data. The squeezing and antisqueezing levels are estimated from the inflection points of the probability density functions (PDFs) of quadrature variances, which are obtained from the Gaussian kernel density estimates (KDEs) of the histograms. The squeezing and antisqueezing level estimates correspond to the locations of the peak slopes at the left (right) edges of the PDF, respectively. In particular, the quadrature variances for the squeezing and antisqueezing levels are identified from the peaks in the derivative of the KDEs, which provide a well-defined measure of the edges of the quadrature variance distribution. The same estimation procedure applied to the vacuum data yields the standard deviation in the vacuum sample variance (shot noise level). Error bars are obtained from the propagation of the vacuum standard deviation on the squeezing and antisqueezing level estimates.

\subsection{Measurement characterization}
For each quantum measurement in the reported experiments, a classical measurement is also taken to characterize the system. The classical measurements are taken using the same photonic and electronic hardware chain as the quantum measurements to ensure consistency. For the experiment in Fig. \ref{fig:fig3}b, a classical multipixel image is taken by sending a coherent state as signal while the LO phase is ramped at 5 MHz. The 5 MHz tone from each channel is digitized by the imaging readout, and its corresponding amplitude is measured. For the experiments in Figs. \ref{fig:fig4}b,e and f, a coherent state is sent as signal while the LO phase is ramped at 5 MHz. The 5 MHz tone at the output of the power combiner is measured on the ESA for each measurement setting. For each channel combination in Fig. \ref{fig:fig4}b, an SNR is calculated by taking the ratio of this signal power to the respective vacuum noise acquired from the squeezed light measurement.

\subsection{Wigner function calculation}
For the calculations of the Wigner functions in Fig. \ref{fig:fig3}b, the experimental squeezing parameter $r=1.95$ is plugged into the Wigner function $W(X,P, r,\theta, \eta)$ of a squeezed vacuum state \cite{leonhardt1993realistic}, setting $\theta = 0$ and $\eta = 1$ to obtain the Wigner function at the source. The Wigner function for each pixel mode is obtained by plugging the squeezing parameter, phase, and geometric efficiency into  $W(X,P, r,\theta, \eta)$. The phases are estimated from a sinusoidal fit to the sample variances of each channel over a region of the data where the phase modulation was approximately uniform. From the squeezing parameter, the effective efficiency of each channel is estimated using,
\begin{align}
    \eta = \frac{(A-1)\exp(2r)}{(\exp(2r)-1)(A+\exp(2r))}, \label{eq:eta_p2p}
\end{align}
where $A = \Delta X_+^2/\Delta X_-^2$ is the ratio of the antisqueezing ($\Delta X_+^2$) to squeezing ($\Delta X_-^2$) levels. The geometric efficiencies of the pixels are calculated from the effective efficiencies of the channels divided by their total sum.

\subsection{Theoretical modeling}
The theoretical models in Fig. \ref{fig:fig4} are constructed from the classical data of the measurement characterizations using,
\begin{align}
    \Delta X_{\pm}^2 = \eta e^{\pm2 r} + 1-\eta, \label{eq:models}
\end{align}
where $\Delta X_\pm^2$ are the squeezing ($-$) and antisqueezing ($+$) levels and $r$ is the squeezing parameter, and $\eta$ is the effective efficiency of the system. 

For Fig. \ref{fig:fig4}b, the model is obtained from Eq. $\ref{eq:models}$ with $\eta \propto \text{SNR}$. A least-squares fit is performed by taking the proportionality constant ($\eta_c$) to the classical SNR data as the only free parameter, with the squeezing parameter bounded in the range $r = 0.748\pm0.019$.  Using SNR data normalized to its peak value, we obtain optimal parameters of $\eta_c = 0.021$ and $r = 0.761$. 

For Fig. \ref{fig:fig4}c, the model is obtained from Eq. $\ref{eq:models}$ with $r = \mu\sqrt{P}$, where $P$ is the squeezer pump power and $\mu$ is the SPDC efficiency, and a least-squares fit is performed taking the $\mu$ and $\eta$ as free parameters. The optimal parameters are reported in the main text. 

For Fig. \ref{fig:fig4}e, the models are obtained from Eq. $\ref{eq:models}$ and $\eta$ proportional to classical beamwidth data for 8 and 32 channels combined. For each data set, a least-squares fit is performed taking the proportionality constant ($\eta_c^{(n)}$) to the classical beamwidth data as the only free parameter, with the squeezing parameter bounded in the range $r=0.607\pm0.015$. Using beamwidth data normalized to their peak powers, we obtain optimal parameters of $\eta_{c}^{(8)} = 0.019$, $\eta_{c}^{(32)} = 0.014$, and $r = 0.611$. The 8 and 32 channel beamwidths are characterized directly from the squeezed light data by extracting the effective efficiencies using Eq. \ref{eq:eta_p2p}. With linear interpolation, angles corresponding to 0.5 effective efficiency are found to calculate the beamwidths.

For Fig. \ref{fig:fig4}f, the models are obtained from Eq. $\ref{eq:models}$ and $\eta$ proportional to the classical radiation pattern of a single antenna. For each data set, a least-squares fit is performed taking the proportionality constant ($\eta_{c}^{(n)}$) to the classical radiation pattern as the only free parameter, with the squeezing parameter bounded in the range $r=0.865 \pm 0.043$. Using the radiation pattern data normalized to its peak power, we find optimal parameters of $\eta_{c}^{(8)} = 0.017$, $\eta_{c}^{(32)} = 0.015$, and $r = 0.908$.  The 8 and 32 channel FoVs are characterized directly from the squeezed light data in the same way as beamwidth characterization using Eq. \ref{eq:eta_p2p}.

The squeezing parameters for the models are obtained from independent characterizations of the sources.

\subsection{Cluster state generation} 
Cluster states of up to eight modes have been demonstrated with bulk multipixel homodyne detection systems by programming virtual optical networks in digital post-processing  \cite{armstrong2012programmable}. The virtual networks mix different spatial regions in a beam of light to match the detection basis to an entangled spatial mode basis. 
This method of entanglement generation allows for highly compact and versatile implementations of Gaussian quantum computation in the measurement-based model \cite{ferrini2013compact}, which can be scaled to a higher number of modes by interfacing quantum PICs like the QPA chip with special-purpose RF or microwave ICs.

With our architecture in Fig. \ref{fig:fig5}a, the overall transformation on the input field can be summarized as,
\begin{align}
    \vec{a}_\text{out} = S (G \oplus G) D U \vec{a}_\text{in},
\end{align}
where $U$ is the free-space change-of-basis unitary mapping the input modes to pixel modes, $D = \text{diag}( e^{i\phi_1},e^{i\phi_2},...)$, 
\begin{align}
    G\oplus G& = \begin{pmatrix}
        1&1&1&1&0&0&0&0\\
        0&0&0&0&1&1&1&1
    \end{pmatrix}
\end{align}
is the gain matrix of the RF power combiners, and
\begin{align}
    S&= \frac{1}{\sqrt{2}}\begin{pmatrix}
        1&i\\
        i&1
    \end{pmatrix}
\end{align}
is the beamsplitter matrix. 
The transformation of $S$ is performed on the digitized quadratures as an emulation of an RF directional coupler, where complex matrix elements are implemented as a $\pi/2$ phase shift. 
For a two-mode Gaussian cluster state generated with $S$, the cluster state correlations are given by,
\begin{align}
    \text{Var}(\hat{X}_3(\theta) - \hat{P}_4(\theta)) &= \text{Var}(\hat{X}_1(\theta)),\\
    \text{Var}(\hat{X}_4(\theta) - \hat{P}_3(\theta)) &= \text{Var}(\hat{X}_2(\theta)),
\end{align}
where $ \text{Var}(\hat{X}_{i}(\theta))$ for $i=1,2$ is given by Eq. \ref{eq:sq_var} for squeezed modes, such that the right-hand side is zero at $\theta = 0$ in the limit of large squeezing parameter and low loss. We note that in our experiment, the inseparability given by Eq. \ref{eq:insep} has a lower bound of 0.5 since the squeezed light was generated in a single mode. This lower bound can be overcome by transmitting multiple squeezed modes to the chip, allowing for the generation of large cluster states up to 32 modes.

\subsection{Chip losses}
On-chip losses consist of 3.78 dB loss from simulated antenna insertion loss, 0.321 dB loss from waveguide propagation loss, and 1.52 dB loss from photodiode quantum efficiency. This results in a total expected on-chip loss of 5.62 dB. In addition to on-chip losses, there is also the geometric loss due to the mode mismatch between the aperture and the collimated beam and the insertion loss of the collimator. The on-chip losses are verified experimentally by sending 200 $\mathrm{\mu m}$ collimated beam to the chip aperture after setting all QRXs to the unbalanced (100:0) configuration and summing all QRX currents. For 0.452 mW input power, the output current is 0.0615 $\mathrm{\mu A}$, resulting in an insertion loss of 8.66 dB. For a 200 $\mathrm{\mu m}$ collimated beam, the geometric loss is 1.14 dB, the insertion loss of the collimator is 0.8 dB, and the insertion loss of the connectors is expected to be \textless1 dB. De-embedding these losses from the measurement, the on-chip losses are measured to be 5.72 dB, which agrees well with the 5.62 dB expected loss.
\end{document}